# High electron mobility single-crystalline ZnSnN$_2$ on ZnO (0001) substrates

D. Gogova[1,*], V. S. Olsen[1], C. Bazioti[1], I.-H. Lee[1,2], Ø. Prytz[1], L. Vines[1], and A. Yu. Kuznetsov[1]

[1]Department of Physics, Center for Materials Science and Nanotechnology, University of Oslo, PO Box 1048, Blindern, NO-0316 Oslo, Norway

[2]Department of Materials Science & Engineering, Korea University, Seoul, Republic of Korea

Abstract

Making a systematic effort, we have developed a single-crystalline ZnSnN$_2$ on ZnO (0001) by reactive magnetron co-sputtering. Epitaxial growth was achieved at 350°C by co-sputtering from metal targets in nitrogen atmosphere, and confirmed by transmission electron microscopy (TEM) measurements. TEM verified that the layers are single-crystalline of hexagonal phase, exhibiting epitaxial relationship with the substrate described by: $[11\text{-}20]_{ZnSnN2}//[11\text{-}20]_{ZnO}$ and $[0001]_{ZnSnN2}//[0001]_{ZnO}$. The screw-type threading dislocations originating from the ZnSnN$_2$/ZnO interface were identified as dominant extended defects. More specifically, we report a pioneering measurement of the dislocation density in this material of 1.5 x 10$^{11}$ cm$^{-2}$. Even though, there is no literature data for direct comparison, such values are typical of heteroepitaxial growth of III-nitride layers without applying defect density reduction strategies. The films demonstrated high electron mobility of 39 cm$^2$V$^{-1}$s$^{-1}$ and 63 cm$^2$V$^{-1}$s$^{-1}$ for stoichiometric and Zn-rich layers, respectively, while the electron carrier density remained in the low 10$^{19}$ cm$^{-3}$ range as determined by the Hall effect measurements at room temperature. The optical bandgaps of 1.86 eV and 1.72 eV were determined for the stoichiometric and Zn-rich samples, respectively. As such, we conclude that ZnSnN$_2$ is an earth-abundant, environmentally-friendly semiconductor and is a promising candidate for cost efficient components in electronics and photonics.

## 1. Introduction

During the last decade, III-nitrides revolutionized solid state lighting and power electronics as was foreseen already in the 1990s [1]. Expanding the nitride semiconductor family by including the Zn-IV-$N_2$ heterovalent ternary nitrides offers a new opportunity for device design that may help to overcome some of the limitations of the binary nitrides. The class of Zn-IV-nitride semiconductors is closely related to the group III-N semiconductors in terms of physical properties, i.e., their bandgaps can be tuned from the ultraviolet to the infrared range of the spectrum [2-4]. The advantages of Zn-IV-$N_2$ in comparison to the group III-nitrides are: (i) composed of Earth-abundant elements, i.e., Zn-IV-$N_2$ are cost-effective in comparison to III-Ns; (ii) potential to avoid the phase separation in the whole alloys composition range, unlike e.g. InGaN with high In content, and (iii) lower deposition temperatures, i.e., more environmentally-friendly materials than the III-Ns.

Zinc tin nitride ($ZnSnN_2$) belongs to the Zn-IV-$N_2$ group and has many beneficial properties, including the richer Earth abundance, better recyclability of Zn and Sn than that of In and Ga, no toxicity, and a direct band gap of 1.0 - 2.0 eV. Meanwhile, $ZnSnN_2$ possesses a large absorption coefficient ranging from the ultraviolet through visible light and into the near-infrared, comparable to some typical photovoltaic materials such as GaAs, CdTe, and InP ($\sim 10^5$) [5]. Importantly, $ZnSnN_2$ may potentially demonstrate reasonably high carrier mobility. Due to these favorable properties, $ZnSnN_2$ has a potential to be applied in photovoltaics, optoelectronics, photocatalysis, etc. Furthermore, it is a relatively unexplored member of the Zn-IV-$N_2$ semiconductor family (first experimental report on $ZnSnN_2$ appeared in 2013 [6]) and data on the Zn-Sn-$N_2$ alloys remain scarce and sometimes even contradictory. For example, the crystallographic structure and the optical band gap of the material are still not unambiguously defined.

The very different melting points of Zn and Sn combined with the low decomposition temperature of $ZnSnN_2$ (~ 400°C at one atmosphere [7]) make its bulk growth rather difficult. Therefore, $ZnSnN_2$ is synthesized in thin film forms by MBE, and DC or RF sputtering [2, 6, 8 -11]. So far there is no publication in the literature on CVD growth of $ZnSnN_2$, a process which is close to thermodynamic equilibrium [12,13], partly due to issues with the CVD precursors having appropriate vapor pressures. Meanwhile, MBE and sputtered $ZnSnN_2$ films could exhibit multiple crystallographic orientations and are strained due to the contact with hetero-substrates because of the lattice and thermal expansion coefficients mismatch. Reasonably good single crystalline $ZnSnN_2$ films have been reported on cubic yttria-stabilized zirconia (111) [11,14], polar and nonpolar (1-102) sapphire [11], $GaN/Al_2O_3$(0001) [6] and on ZnO-buffered sapphire [15]. However, mastering of the synthesis is far from being mature. In particular, it is unknown whether the use of the bulk ZnO substrates may have an impact on the film quality, e.g. because of a better thermal expansion match comparing to sapphire.

In our work, reactive magnetron sputtering was selected to synthesize the samples due to its high flexibility in alloying elements with rather different physical and chemical properties, cost-effectiveness, capability to achieve reasonably good crystalline quality, combined with high growth rates and uniformity on large areas, i.e., compatibility with the industrial requirements. Making a systematic study, we have selected a variety of different substrates and tested a broad range of the growth parameters. As a result of the growth optimization, we demonstrate a record high electron mobility ZnSnN$_2$ single-crystalline films epitaxially grown on lattice parameter matched ZnO (0001) bulk substrates and measure its bandgap both for the stoichiometric and Zn-rich compositions.

## 1. Experimental

Zn$_x$Sn$_{1-x}$N$_2$ (0 ≤ x ≤ 1) films with a wide compositional x-range were deposited by magnetron co-sputtering of metal Zn (99.995 vol.%) and Sn (99.995 vol.%) targets in N$_2$-containing atmosphere employing a Moorfield RF Magnetron Sputter equipment. The target diameter was 75 mm. They were inclined at 45° to the substrate normal and the distance between the targets and the substrate holder was fixed to 100 mm. In most experiments pure N$_2$ (99.999%) was used as a sputter gas, in some – a small amount of argon (99.999%) was added to decrease the nitrogen fraction in the synthesized films. For calibration reasons, binary Zn-N and Sn-N were sputtered to optimize the deposition conditions and to determine the corresponding deposition rates. The chamber base pressure was ~ 6.7x10$^{-5}$ Pa.

The substrates with lateral sizes 10x10 or 5x5 mm were cleaned in an ultrasonic bath as follows: 5 min in acetone, 5 min in propanol, 5 min in deionized water and finally dried in a N$_2$-flow of high purity. In order to have a basis for comparison, the growth was performed on a variety of amorphous and single-crystalline substrates, e.g. quartz, soda-lime glass, Si(100), glassy carbon (Sigradur R), c-, r- and m-plane sapphire, ZnO(0001)±0.5° (hydrothermally grown, double-side polished, MTI Co.), and home-made ZnO/Al$_2$O$_3$(0001) since the Zn-Sn-N$_2$ has order-dependent properties [16]. The availability of the ZnSnN$_2$ films on different substrates also allowed us to choose appropriate samples for the characterization methods. The deposition temperature was varied from room temperature to 550°C with a step of 50°C, and the sample stage was rotated at 11 rpm to ensure the films uniformity. Heating up and cooling down in a nitrogen-containing atmosphere was employed. The ZnSnN$_2$ films prepared were nominally-undoped. Notably, the sapphire and ZnO substrates were pre-annealed at 1050°C for 1 h in oxygen-containing atmosphere to prepare an epi-ready surface.

The structural quality and phase were studied by x-ray diffraction (XRD) employing a Bruker AXS D8 diffractometer. The X-ray source was Cu K$_{\alpha1}$ ($\lambda$ = 1.5406 Å), and a Ge (220) double bounce monochromator was implemented to filter out the K$_{\alpha2}$ signal ($\lambda$ = 1.5444 Å). The thin film thickness was determined employing a Dektak 8 stylus profilometer and by means of a Scanning Electron Microscope (SEM) JEOL IT-300, while the films composition – by Energy Dispersive X-ray Spectroscopy (EDX) in the SEM on Zn-Sn-N$_2$ samples deposited on oxygen-free substrates: Si(100) and glassy carbon. The EDX impurity detection limit is of 0.5 at.%.

(Scanning) Transmission Electron Microscopy (S)TEM investigations were conducted on a FEI Titan G2 60-300 kV equipped with a CEOS DCOR probe-corrector. Observations were performed at 300 kV with a probe convergence angle of 24 mrad. The camera length was set at 77 mm and simultaneous STEM imaging was conducted with 3 detectors: high-angle annular dark-field (HAADF) (collection angles 98.7-200 mrad), annular dark-field (ADF) (collection angles 21.5-98.7 mrad) and annular bright-field (ABF) (collection angles 10.6 - 21.5 mrad). The resulting spatial resolution achieved was approximately 0.08 nm. Microprobe diffraction patterns were acquired with a 4 nm electron probe size. The samples for STEM with a cross-sectional wedge geometry were prepared by mechanical grinding and polishing (Allied MultiPrep). Final thinning was performed by Ar ion milling with a Fishione Model 1010, and plasma cleaning was applied before the STEM investigations, with a Fishione Model 1020.

Mass spectra and impurity versus depth profiles were obtained by Secondary Ion Mass Spectrometry (SIMS) using a Cameca IMS 7f microanalyzer with 15 keV $Cs^+$ ions as the primary beam. The depth of the sputtered crater was measured by a Veeco Dektak 8 stylus profilometer. Assuming a uniform and time-independent erosion rate, the measured crater depth was used to convert sputtering time to sample depth. On the other hand, the calibration of the SIMS signal intensity into the atomic concentration was not done, because of the absence of the reference samples for these new materials. Notably, since nitrogen is known to be difficult to ionize, it was monitored in a complex with Zn.

The optical properties were deduced from transmittance and reflectance measurements of the samples in the 175-2600 nm range, with a resolution of 0.1 nm, employing a spectrophotometer of type Shimadzu SolidSpe-3700.

The free carrier concentration and mobility data were collected at room temperature by Hall effect measurements, employing a Lakeshore 7604 device in a van der Pauw configuration of the electrodes. To make the data reliable while measuring on the conductive substrates, specifically on n-type ZnO, a current blocking high resistive surface region was made by nitrogen ion implantation [17]. In short, ZnO was implanted with $2\times10^{14}$ $N/cm^3$ (180 keV) and $1\times10^{14}$ $N/cm^3$ (36 keV) at room temperature and subsequently annealed at 1050°C for 1 h in oxygen-containing atmosphere to recover the implantation damage.

## 3. Results and Discussion

Several series of Zn-Sn-$N_2$ thin films were deposited by magnetron co-sputtering on a variety of amorphous and single-crystalline substrates to exploit different structures. Concurrently, as a prime goal, we were targeting the deposition of the epitaxial material. For this purpose, substrates with the best lattice parameters and thermal expansion coefficients matching to $ZnSnN_2$ were selected for the growth optimization and detailed investigations. Specifically, bulk ZnO wafers suited best (JCPDS card No 36-1451, hexagonal: a = 3.252 Å, c = 5.313 Å), providing the lattice mismatch between $ZnSnN_2$ and ZnO of approximately 4%. Moreover, Zn-Sn-$N_2$ thermal expansion matching with conductive n-type ZnO`s was expected to be better that with the insulating substrates, e.g. sapphire, even though the literature data are missing.

Importantly, the growth window for deposition of $ZnSnN_2$ thin films (amorphous and polycrystalline) is relatively wide, i.e., the deposition temperature can be varied from room temperature to about 550°C. In particular, our growth optimization data are in agreement with that reported by Kawamura et al. [7] for the $ZnSnN_2$ decomposition. For example, we observed that the growth rate drops from 0.9 nm/min at 350°C to 0.33 nm/min at 450°C. Thus, unlike $ZnGeN_2$ and $ZnSiN_2$ which are high-temperature stable materials [18,19], the lower decomposition temperature of $ZnSnN_2$ limits the synthesis temperature and hinders further crystallinity improvement by elevating the growth temperature. Notably, Cai and co-workers [20] reported 350°C as a suitable substrate temperature to obtain polycrystalline ZnSnN$_2$ thin films on Si(100) and K9 glass, but we managed to grow a single-crystalline material at that temperature using better matching substrates and systematically optimized deposition parameters.

The impact of all deposition parameters during reactive co-sputtering of ZnSnN$_2$ on its structural, compositional, optical and electrical properties will be reported elsewhere. In short, we found 350°C to be the optimal deposition temperature, while the variations in the RF power helped to fine-tune the composition. As a result, in the rest of the paper, we mainly compare two samples - stoichiometric ZnSnN$_2$ and non-stoichiometric - Zn$_{1.60}$Sn$_{0.40}$N$_2$, both grown at 350°C, N$_2$ flow-rate of 24 sccm, and sputtering pressure of 0.08 Pa. Upon the optimization, for the stoichiometric and Zn-rich sample, the Zn to Sn power ratios on the targets were 28:78 W and 78:28 W, respectively. Notably, the crystallinity of both samples was comparable.

### 3.1. Compositional study

EDX measurements performed in the SEM microscope confirmed the stoichiometric composition of the samples grown at Zn:Sn=28:78 W, i.e., the nitrogen content of 50 at.%, as well as zinc and tin contents of 25 at.% each. No other impurities were detected in the epitaxial layers with the sensitivity of the EDX. For the Zn-rich sample the Zn and Sn content were determined as: 40 at.% and 10 at.%, respectively.

Mass spectra were collected in a broader range using SIMS, revealing mainly oxygen and hydrogen related impurity signals in the films. Further, SIMS depth profiling of ZnSnN$_2$ layers grown on ZnO was employed to assess depth uniformity, where Fig.1 display the SIMS intensity versus depth obtained for the matrix elements as well as hydrogen and oxygen. Indeed, the results (Fig. 1) reveal a uniform distribution of both matrix elements and impurities.

From EDX incorporated in the SEM, no oxygen signal was observed, and hence below the detection limit of our system (approximately 0.5 at%). However, oxygen is present in the films to a concentration detectable by SIMS (see Fig. 1) and may be attributed to the residual contaminations due to the relatively high chamber base. The same reason holds for explanation of the hydrogen contamination.

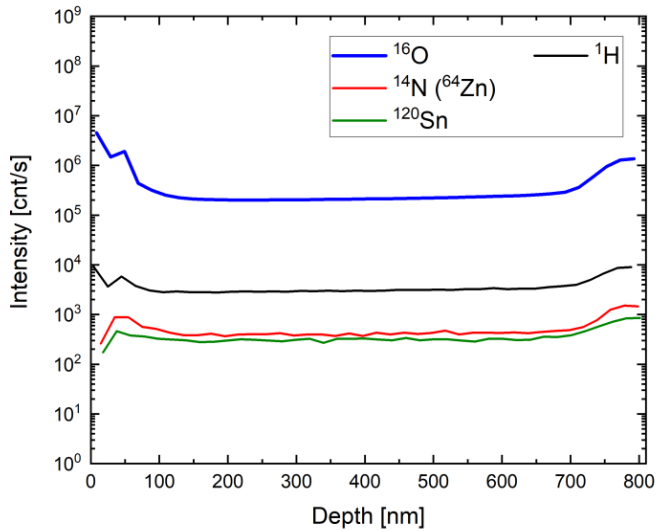

Fig. 1 Example of SIMS vs depth profiles acquired employing Cs$^+$ as primary ions. The nitrogen related signal (red line) was evaluated using mass corresponding to a complex with zinc, hence, the signal represents the combined concentration of both N and Zn. Note that the ionization efficiency of oxygen (blue) are considerably higher than that of the other measured elements, and therefore results in a substantially stronger signal intensity (in counts/s). Hydrogen (black) and tin (green) were also recorded.

From the $^{16}$O signal, it is furthermore observed indications of surface oxidation, with the increase in $^{16}$O signal close to the layer surface, which probably stems from storing the films at ambient conditions. Furthermore from Fig. 1, the detection of a hydrogen related signal is observed, also incorporated during growth due to residual gas in the growth chamber.

### 3.2 Structural properties

A θ-2θ-scan, typical of stoichiometric ZnSnN$_2$ grown on ZnO (0001) at 350°C, is illustrated in Fig. 2. The intense and narrow peak which appears at 2θ = 32.65°, can be assigned to the (0002) Bragg reflection and this with less intensity at 2θ = 68.35° to the (0004) reflection of the wurtzite ZnSnN$_2$ [6,20].

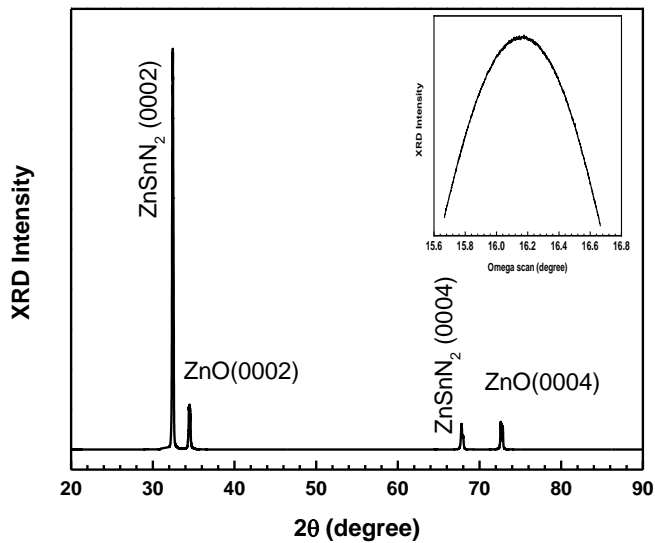

Fig. 2 XRD θ-2θ-scan of ZnSnN$_2$ epitaxial layer grown on ZnO(0001) and the Rocking curve of the (0002) peak of ZnSnN$_2$ (in the inset).

The θ-2θ-scan reveals two additional peaks – the (0002) reflection of wurtzite ZnO is situated at 34.331°, and the (0004) one at 72.675°. Thus, there exists a well-defined epitaxial relationship between the ZnO substrate and the $ZnSnN_2$ epilayer, i.e., we have grown an epitaxial material with the surface orientation of the ZnO substrate and hexagonal crystallographic structure, confirmed by (S)TEM (see the data and discussion below).

According to the literature data, $ZnSnN_2$ may crystallize in several structures: orthorhombic Pmc21, orthorhombic Pna21, hexagonal P63mc, or monoclinic [2-4, 6, 8-11]. Theoretically, the most stable space group for $ZnSnN_2$ is predicted to be the orthorhombic Pna21 [6]. Nevertheless, no additional peaks, at ~15° and at ~20-25°, which can be assigned to the orthorhombic phases have been observed in the θ-2θ--scans of our stoichiometric layers. The full width at half maximum of the (0002) Rocking curve (see the inset of Fig. 2) was determined to be 0.8°. Such good crystalline quality has been reported for MBE grown epitaxial $ZnSnN_2$ material on $ZnO/Al_2O_3$(0001) only – 0.52 and 0.71° at 450 and 550°C [15]. However, our layers demonstrate significantly higher carrier mobility and a flexibility of the bandgap tuning (as will be discussed in Sec. 3.3 and Sec. 3.4).

For the structural investigation of the epitaxial layers on the nanoscale, (S)TEM was employed. In Fig. 3(a) an ABF-STEM image illustrates the overall morphology of the $ZnSnN_2$ layer, with an average thickness ~200 nm. The Selected Area Electron Diffraction (SAED) pattern (inset) verified that the layers are single crystals of hexagonal phase, exhibiting a good heteroepitaxial orientation relationship with the ZnO substrate described by: $[11-20]_{ZnSnN2}//[11-20]_{ZnO}$ and $[0001]_{ZnSnN2}//[0001]_{ZnO}$. The lattice parameters were extracted by measurements on SAED patterns (that were calibrated using the substrate as a calibration standard) and found to be equal to: $a = 0.339 \pm 0.005\ nm, c = 0.546 \pm 0.009\ nm$. The resulting mismatch with the ZnO substrate is 4.2% (using a= 0.3249 nm and c= 0.5206 nm for ZnO).

The dominant extended defects in ZnSnN$_2$ epitaxial layers are threading dislocations (TDs) emanating from the ZnSnN$_2$/ZnO interface. TEM imaging under two-beam conditions using *g0002* and *g1-100* (not shown here) revealed that the TDs are mainly c-type (screw TDs) and the TD-density was 1.5 x 10$^{11}$ cm$^{-2}$. More specifically, we report a pioneering measurement of the dislocation density in this material of 1.5 x 10$^{11}$ cm$^{-2}$. Even though, there is no literature data for direct comparison, such values are typical of heteroepitaxial growth of III-nitride layers [21] without defect density reduction strategies [22-24]. Nitrides are known as defect-tolerant semiconductors, i.e., the high defect densities do not disturb considerably their functional properties [21].

Formation of an interfacial zone of ~15 nm was detected inside the layer. Microprobe diffraction patterns in STEM-mode were acquired, scanning different regions with the electron-beam having a probe size ~ 4 nm. Figure 3(b) shows ADF-STEM images of the ZnSnN$_2$/ZnO heterostructure, with the red cross marking the position of the electron beam and Figure 3(c) shows the corresponding microprobe diffraction patterns. No structural difference was observed between the interfacial area and the rest of the epitaxial layer. Further investigations are pending in order to understand the properties and the formation mechanism of this area.

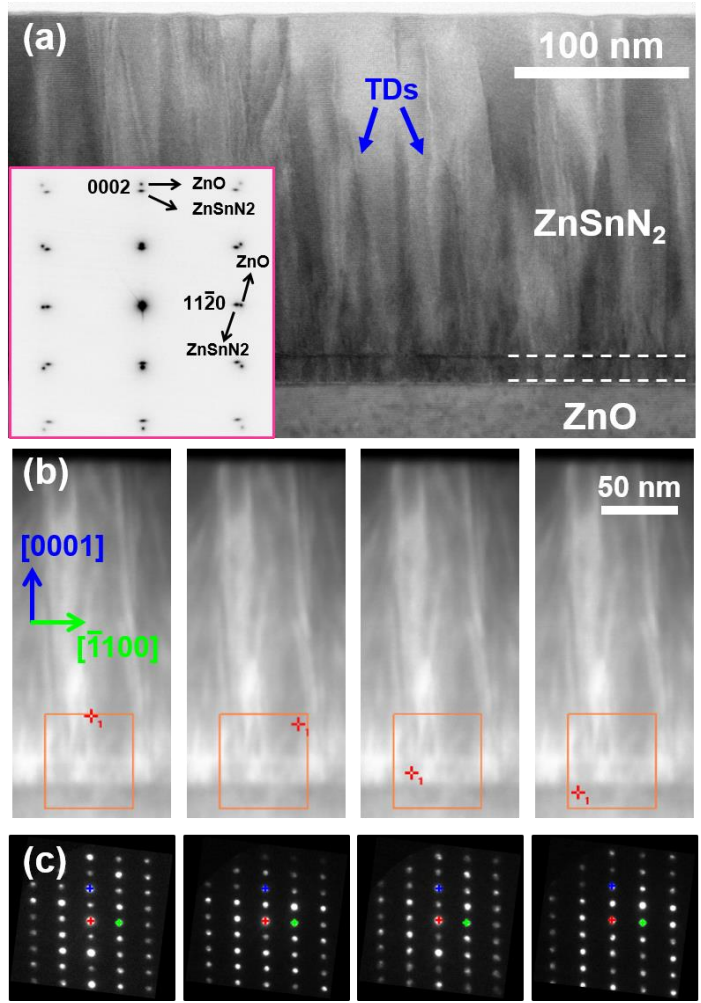

Fig. 3 (a) ABF-STEM image illustrating the overall morphology of the $ZnSnN_2$ layer grown along the [0001] direction on c-plane ZnO substrate. Threading dislocations of screw-type emanating from the $ZnSnN_2$/ZnO interface were the dominant defects, while a self-formed interfacial zone of ~15 nm was detected (dashed lines). Selected Area Electron Diffraction Pattern (inset) verified that the films are single crystals of hexagonal phase, exhibiting a very good heteroepitaxial orientation relationship with the substrate. (b) ADF-STEM images and (c) the corresponding microprobe diffraction patterns acquired from the area marked by the red cross. No structural changes were detected between the interfacial zone and the rest of the film. (red: direct beam, blue: 0002, green: 1-100).

## 3.3 Electrical properties

The electrical properties of two typical Zn-Sn-N$_2$ samples: - stoichiometric (ZnSnN$_2$ and non-stoichiometric, Zn-rich with Zn$_{1.60}$Sn$_{0.40}$N$_2$ composition were studied by Hall effect measurements at room temperature and the results are summarized in Table 1. All Zn-Sn-N$_2$ layers prepared were nominally-undoped, however, the carrier concentrations as measured by Hall effect are relatively high. Both intrinsic defects, e.g. the Sn$_{Zn}$ antisites, and impurity defects such as O$_N$, H$_N$ and H$_i$, may potentially be responsible for the high carrier concentration. It can be noted, that unintentional doping is a common issue in the development of new semiconductors, including Zn-Sn-N$_2$ alloys. For example, oxygen is a well-known donor in all nitrides since the formation energy of oxygen on a nitrogen site (O$_N$) in the lattice is lower than of that of the nitrogen vacancy (V$_N$) and therefore more likely to form.

Nevertheless, in spite of the unintentional doping issues, we have grown Zn-Sn-N$_2$ epitaxial layers with the highest up to date reported electron mobility: 63 cm$^2$V$^{-1}$s$^{-1}$ and 39 cm$^2$V$^{-1}$s$^{-1}$ for Zn-rich and stoichiometric material, respectively. For comparison, 7 cm$^2$V$^{-1}$s$^{-1}$ has been reported for amorphous ZnSnN$_2$ thin films [25]; Ref. 18 demonstrated epitaxial ZnSnN$_2$ prepared at a work pressure of 7.0 Pa with an electron density of 6.72×10$^{19}$ cm$^{-3}$ and a mobility of 24.3 cm$^2$V$^{-1}$s$^{-1}$. Our record high mobility values may be attributed to the higher crystallinity of the material as a result of using better lattice mismatched bulk ZnO substrate. Moreover, in comparison with ZnSnN$_2$ epitaxially grown on sapphire or on ZnO-buffered sapphire substrate [15], the defect concentration in our samples may be lower due to better thermal expansion matching. Notably, the Zn-rich growth conditions result in a 25% decrease in the free-carrier concentration (see Table 1), likely due to the reduced amount of native donor-type (Sn on $Zn^{2+}$ site) defects [14]. By depositing Zn-Sn-$N_2$ films in higher vacuum and using Zn/(Zn+Sn)=0.72, Wang et al. [14] have achieved more pronounced Zn-excess effects, reaching $2.7 \times 10^{17}$ electrons/cm$^3$ at room temperature, however, reported mobility below 3 cm$^2$V$^{-1}$s$^{-1}$. In this sense, our result of 63 cm$^2$V$^{-1}$s$^{-1}$ in the non-stoichiometric material – see Table 1 – is remarkable, and may be interpreted in terms of a significant suppression of the concentration of the scattering centres.

Our attempts to further decrease the background concentration by passivation during growth and/or ex-situ annealing, i.e., to decrease the unintentional donor formation in ZnSnN$_2$ using doping control approaches employed in III-Ns are still in progress.

Table 1 Electrical properties of stoichiometric (ZnSnN$_2$) and non-stoichiometric (Zn$_{1.60}$Sn$_{0.40}$N$_2$) epitaxial layers at 300 K.

| Sample composition | Thickness (nm) | Cation ratio Zn/(Zn+Sn) | Mobility (cm$^2$V$^{-1}$s$^{-1}$) | Electron concentration (cm$^{-3}$) | Resistivity (Ω.cm) |
| --- | --- | --- | --- | --- | --- |
| ZnSnN$_2$ | 350 | 0.5 | 38 | $-2.11 \times 10^{19}$ | $7.63 \times 10^{-3}$ |
| Zn$_{1.60}$Sn$_{0.40}$N$_2$ | 240 | 0.8 | 63 | $-1.61 \times 10^{19}$ | $6.14 \times 10^{-3}$ |

### 3.4 Optical properties

The optical bandgap of the Zn-Sn-$N_2$ epitaxial layers was determined from transmission and reflection measurements. The absorption coefficients were calculated using the relation: $T = (1 - R)^2 \exp(-\alpha^* d)$, where $d$ is the layer thickness.

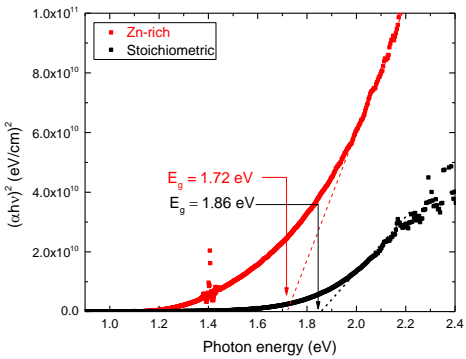

Fig. 4 Tauc plot for direct, allowed transitions of the stoichiometric (ZnSnN$_2$) and the Zn-rich - (Zn$_{1.60}$Sn$_{0.40}$N$_2$) layer.

Fig. 4 depicts the Tauc plot of the two typical sample compositions: ZnSnN$_2$ and Zn$_{1.60}$Sn$_{0.40}$N$_2$. Both Zn-Sn-N$_2$ epitaxial layers demonstrate a very strong absorption, so that significant noise emerges at high energies as a result of the extremely low layers transmittance. The points at 1.4 eV are artefacts due to the detector change. The optical bandgap ($E_g$) values were determined as: 1.86 eV for the stoichiometric sample and 1.72 eV for the Zn-rich Zn-Sn-N$_2$ layer. With the increase of the Zn:Sn ratio, the absorption edge shifts towards lower energies in accordance with the literature data for amorphous and microcrystalline Zn-Sn-N$_2$ thin films [25]. The bandgap values of our material are larger than the theoretically calculated 1.42 eV at 0 K [6]. One of the possible reasons is that the absorption edge is pushed to higher energies because of the so-called Burstein–Moss shift in semiconductors [26]. Notably, the difference in the bandgaps for the stoichiometric and Zn-rich layers indicated an option for the corresponding quantum well design in multiple stacks of the layers, similarly to that in III-N materials.

**Conclusions:**

Zn-IV-N$_2$ alloys comprise a promising class of semiconductors since its bandgap energy could cover the solar spectrum and may compete to replace rather expensive III-Ns alloys in many applications. As a result of a systematic effort, we have developed the single-crystalline ZnSnN$_2$ material on ZnO (0001) by reactive magnetron co-sputtering. Even though the orthorhombic structure has been theoretically predicted to be energetically favourable in literature, our TEM results confirmed that the wurtzite crystalline structure becomes dominating in the films grown on appropriate substrates using optimized growth conditions. The lattice parameters of the hexagonal stoichiometric ZnSnN$_2$ were determined as: $a = 0.339 \pm 0.005 \; nm, c = 0.546 \pm 0.009 \; nm$. The screw-type threading dislocations originating from the ZnSnN$_2$/ZnO interface were identified as dominant extended defects. More specifically, we report a pioneering measurement of the dislocation density in this material of 1.5 x 10$^{11}$ cm$^{-2}$. Even though, there are no literature data for direct comparison, such values are typical of heteroepitaxial growth of III-nitride layers without applying defect density reduction strategies. The films demonstrated high electron mobility of 39 cm$^2$V$^{-1}$s$^{-1}$ and 63 cm$^2$V$^{-1}$s$^{-1}$ for stoichiometric and Zn-rich layers, respectively, while the electron carrier density remained in the low 10$^{19}$ cm$^{-3}$ range as determined by the Hall effect measurements at room temperature. The optical bandgaps of 1.86 eV and 1.72 eV were determined for the stoichiometric and Zn-rich samples, respectively, providing an option for the corresponding quantum well design in multiple stacks of the layers, similarly to that in III-Ns. As such, we conclude that ZnSnN$_2$ is the earth-abundant, environmentally-friendly semiconductor and is a promising candidate for cost efficient components in electronics and photonics as well as for the emerging next generation PV's solar cells.

**Acknowledgements:**

This work has been funded by the Research Council of Norway (RCN), which is acknowledged for support to the SALIENT project (239895/F20). RCN is also acknowledged for the support to the Norwegian Micro- and Nano fabrication Facility (NorFab), project number 245963/F50 and the Norwegian Center for Transmission Electron Microscopy (NORTEM), project number 197405/F50. IHL acknowledges the financial support for his sabbatical stay at the Department of Physics and the Center for Material Sciences and Nanotechnology at the University of Oslo.

**Conflicts of interest**
There are no conflicts to declare.